# Tracking the resonant frequency of a micromechanical resonator using phononic frequency combs

Adarsh Ganesan[1] and Ashwin Seshia[1]

[1.] Nanoscience Centre, University of Cambridge, Cambridge, UK

**Micro and nanomechanical resonators have been extensively researched in recent decades for applications to time and frequency references, as well as highly sensitive sensors. Conventionally, the operation of these resonant sensors is practically implemented using a feedback oscillator to dynamically track variations in the resonant frequency. However, this approach places limitations on the frequency stability of the output response, particularly owing to near-carrier phase noise, potentially limiting measurements over long integration times. Here, in this paper, utilizing the recent experimental demonstration of phononic frequency combs, we demonstrate an alternative resonant tracking approach with the potential to provide significant improvements in near-carrier phase noise and long-term stability. In addition, we also showcase comb dynamics mediated resonant frequency modulation which indirectly points to the possible control of inevitable noise processes including thermomechanical fluctuations. This resonant tracking approach may also have general applicability to a number of other physical oscillators.**

The tracking of resonant frequency associated with physical oscillators is fundamental to timing [1-4] and sensing applications [5-8]. Such oscillators include physical systems based on atomic transitions [1,5], quartz crystals [2,6] and micro-and nanomechanical resonators [3-4,7-8]. In recent decades, micro- and nanomechanical resonators have been intensively researched, both for timing and frequency reference applications, as well as to the design of highly sensitive detectors. To implement micro- and nanomechanical oscillator, a feedback configuration is usually employed [9]. Here, white noise injected in the feedback path is shaped by the Lorentzian response of the resonator. The Leeson equation is commonly employed to model the phase noise of such an oscillator [10]. This model addresses noise shaping by the linear resonator transfer function and it can be seen that the phase noise near-carrier assumes a $\frac{1}{f^2}$ form with a roll-over to a flat noise floor determined by the 3-dB bandwidth of the resonator, $f_0/2Q$. The resulting frequency fluctuations captured by an Allan Deviation plot demonstrate a $1/\tau$ relationship between the Allan Deviation and integration time, $\tau$, for frequency fluctuations dominated by thermal noise. In practice, nonlinearities and higher-order noise and drift processes are operative in micro- and nanomechanical resonators limiting frequency stabilities over longer time scales [11-13]. Nonetheless, thermal noise shaping within the resonator feedback loop often dominates oscillator frequency stabilities over short-to-medium integration times with noise sources comprising both mechanical and electrical sources. In recent years, approaches to specifically engineer nonlinearities in a feedback oscillator or coupled oscillator configuration have yielded striking improvements in phase noise and frequency stability of micro/nanomechanical resonator oscillators [14-19].

Here in this paper, using a recently established phononic frequency comb pathway, we demonstrate an alternative approach towards micro/nanomechanical resonant frequency tracking with benefits to frequency stabilities observed relative to the standard feedback configuration. Phononic frequency combs are produced via nonlinear interactions between a driven phonon mode and one or more additional parametrically excited modes [20-21].

The Fermi-Pasta-Ulam (FPU) framework has been previously employed as a basis for describing phononic frequency comb formation [20-21]. A specific system of three coupled modes is looked at in the context of this work, though the appearance of phononic frequency combs in a system of $N$-coupled modes is also possible [22].

$$\ddot{Q}_i = -\omega_i^2 Q_i - 2\zeta_i \omega_i \dot{Q}_i + \sum_{\tau_1=1}^{3}\sum_{\tau_2=1}^{3} \alpha_{\tau_1\tau_2} Q_{\tau_1} Q_{\tau_2} + \sum_{\tau_1=1}^{3}\sum_{\tau_2=1}^{3}\sum_{\tau_3=1}^{3} \beta_{\tau_1\tau_2\tau_3} Q_{\tau_1} Q_{\tau_2} Q_{\tau_3} + P\cos(\omega_d t); \quad i = 1,2,3 \quad (1)$$

Based on this dynamics, for $\omega_d \cong \omega_1 \cong (\omega_2 + \omega_3)$, the excitation of tones $\omega_x \cong \omega_2$ and $\omega_y \cong \omega_3$ which satisfies the condition $\omega_x + \omega_y = \omega_d$ may be expected at high-enough values of the drive level, $P$. This is through the process of 3-mode parametric resonance [23]. However, recent work [24] demonstrated the possibility for the formation of phononic frequency combs using the very same dynamics (eq. (1)) [20]. The spectral lines $\omega_1 \pm n(\omega_d - \omega_1); \omega_x \pm n(\omega_d - \omega_1); \omega_y \pm n(\omega_d - \omega_1)$ are observed at the outset of the formation in this form of frequency combs. Specifically, we now consider the case where the signal levels associated with the spectral lines $\omega_x \pm n(\omega_d - \omega_1); \omega_y \pm n(\omega_d - \omega_1)$ are smaller than those corresponding to $\omega_1 \pm n(\omega_d - \omega_1)$. Hence, the solution of the corresponding frequency comb can be approximately expressed as $\sum_p A_p \cos\left((\omega_1 + p(\omega_d - \omega_1))t\right)$. This solution corresponds to a series of equidistant phase-coherent spectral lines and the concomitant temporal signature is a train of pulses corresponding to the primary mode locked to the input drive $\cong \omega_d$. Specifically, when $A_p = A_{-p}$, the response is symmetric about $\omega_1$ and this mathematical manifestation can enable tracking of resonant frequency $\omega_1$ using techniques such as zero-crossing detection. However, in general, $A_p$ need not always be equal to $A_{-p}$. If the tone $\omega_1$ has the highest amplitude among the frequency comb lines i.e. $A_0 > A_{p\neq 0}$, the frequency $\omega_1$ can still be tracked through appropriate detection schemes. As the resonant frequency varies about the regime where the conditions for the comb generation are satisfied [24], the response continues to conform this characteristic nature, and hence an automatic tracking of $\omega_1$ is achieved. As opposed to feedback oscillators, no external gain / phase feedback elements are required, significantly reducing the design complexity as well as the number of noise sources in the loop.

To experimentally demonstrate the concept of resonant tracking utilizing phononic frequency combs, the device previously used to illustrate the formation of phononic frequency combs via three-mode parametric three-wave mixing [24] is again considered. This device is an AlN-on-Si free-free beam structure of dimensions $1100 \times 350 \times 11\ \mu m^3$ as shown in Fig. 2A with Al electrode patterned on the AlN for actuation and sensing. An electrical signal derived from a waveform generator (Agilent 33220A) is fed to this micromechanical device through one split electrode to excite a length extensional mode (Figure 2B); the output signal extracted from the other electrode is probed using a frequency counter (Agilent 53220A). The experiments were carried out under ambient pressure and temperature conditions.

When the drive condition is set to $S_{in}\left(\frac{\omega_d}{2\pi} = 3.8552\ MHz\right) = 10\ dBm$ and the corresponding response of micromechanical resonator is a frequency comb: $\omega_1 + n(\omega_d - \omega_1)$ where $\frac{\omega_1}{2\pi} \cong 3.8475\ MHz$. This is evidenced by the frequency spectrum presented in Fig. 3A. The temporal trait of this response corresponds to an amplitude modulated sinusoid of frequency $\cong \frac{\omega_d}{2\pi}$ (Figure 3B). This signal is conveyed to the frequency counter to count the high-amplitude tone of $\frac{\omega_1}{2\pi}$ at gate time of $100\ ms$ and the corresponding frequency counts are presented in the figure 3C. The Allan deviation of this dataset is presented in the figure 3D. Based on this, the frequency stability can be quantified as $5.901\ ppb$ at $0.1\ s$ integration time for a representative measurement set, representing an improvement relative to feedback oscillators [25-26] based on similar micromechanical resonators operated under the same conditions. Figure 4 evidences the fact that there are two simultaneous processes operative during the period of frequency combs i.e. (i) nonlinear drift (Figure 4B) and (ii) intrinsic random fluctuations (Figure 4C). The intrinsic random frequency fluctuations can arise from thermomechanical noise and other operative noise processes associated with the nonlinear dynamics of phononic frequency comb formation.

We now turn to the physical origins of the nonlinear drift associated with the frequency comb process. We first hypothesize that this drift in the resonant frequency $\frac{\omega_1}{2\pi}$ is influenced by the parametric excitation and backaction associated with the comb process and is not due to the environmental perturbation. If the resonant frequency drift is merely due to the ambient changes, then the resonant frequency $\frac{\omega_1}{2\pi}$ will drift even in the absence of frequency comb excitation. Hence, we designed the following experiment: switching the frequency comb on for 1 minute 40 seconds and then switching it off for 5 minutes repeatedly for 9 cycles to assess reproducibility across several measurements. The figure 5A shows that the resonant frequency $\frac{\omega_1}{2\pi}$ does not change significantly after each long switch off period. Hence, the modulation of $\frac{\omega_1}{2\pi}$ appears to be dominated by frequency comb dynamics rather than from the environmental perturbation. Further, in the concatenated dataset (Figure 5B), we show a baseline curve connecting all of the individual time series associated with switch-on period. This also proves that there exists a well-defined underlying transient drift pattern related to the comb dynamics induced modulation of $\frac{\omega_1}{2\pi}$. This pattern can be conceived as a rail-track (Figure 5C). While the drift initiates from the value of resonant frequency $\frac{\omega_1}{2\pi}$ at the start of the frequency comb process, its temporal progress depends on the dictated drift pattern (Figure 5D).

The nonlinear drift along with the random fluctuations sets the nominal value of $\frac{\omega_1}{2\pi}$ i.e. in the absence of environmental perturbations. Now, if external disturbances are also introduced, the resonant frequency $\frac{\omega_1}{2\pi}$ will shift away and eventually reach the equilibrium i.e. the nominal value. To experimentally validate this nature, air flow is introduced around the micromechanical resonator by blowing on the device several times (Figure 6). Each time the device is cooled due to the resulting air flow, the resonant frequency shifts and then equilibrates after a characteristic timescale, demonstrating the potential for such a device to be utilised in the context of sensing applications.

In summary, we have demonstrated the tracking of resonant frequency associated with a free-free micromechanical beam resonator utilizing the recently established phononic frequency comb pathway. This approach presents an alternative to the conventional feedback oscillator without the associated complexity of the external circuit and gain / phase feedback control, as well as eliminating the noise sources associated with these external elements that are usually defined by active devices e.g. external transistors. Open-loop frequency tracking is also not susceptible to noise up-conversion in the sense of the feedback oscillator where the output is essentially derived through positive feedback and nonlinear shaping of noise in the feedback loop. The parametric amplification and filtering inherent in the comb dynamics [18-19, 27] also provides for the possibility of noise squeezing in these devices. Furthermore, the practical implementation is relatively straightforward requiring an external drive source only (and no other energy source). This initial implementation has already provided nearly a decade improvement in short-term frequency stability as compared to feedback oscillators based on similar resonators operated under ambient conditions. Further, the evidenced responsivity of the resonant frequency output generated via the frequency comb process points to the potential for its use in sensing applications.

The demonstrated phononic frequency combs based resonant tracking approach may also in general be applicable to a number of other physical oscillators and may potentially prove useful than the standard feedback oscillator configuration. Further work following on these promising results is required to allow for first principles device modelling, and the full theoretical description of the interaction between noise and slow drift processes with the comb dynamics.

**Acknowledgements**

Funding from the Cambridge Trusts is gratefully acknowledged.

**References**

[1] N. Hinkley, J. A. Sherman, N. B. Phillips, M. Schioppo, N. D. Lemke, K. Beloy, M. Pizzocaro, C. W. Oates, and A. D. Ludlow, "An atomic clock with $10^{-18}$ instability," *Science,* vol. 341, pp. 1215-1218, 2013.

[2] W. A. Marrison, "The evolution of the quartz crystal clock," *Bell Labs Technical Journal,* vol. 27, pp. 510-588, 1948.

[3] C. T. C. Nguyen, "MEMS technology for timing and frequency control," *IEEE transactions on ultrasonics, ferroelectrics, and frequency control,* vol. 54, 2007.

[4] J. T. M. Van Beek and R. Puers, "A review of MEMS oscillators for frequency reference and timing applications," *Journal of Micromechanics and Microengineering,* vol. 22, p. 013001, 2011.

[5] J. Kitching, S. Knappe, and E. A. Donley, "Atomic sensors - a review," *IEEE Sensors Journal,* vol. 11, pp. 1749-1758, 2011.

[6] G. Z. Sauerbrey, "The use of quartz oscillators for weighing thin layers and for microweighing," *Z. Phys.,* vol. 155, pp. 206-222, 1955.

[7] R. T. Howe, "Resonant microsensors," in *Proc. 4th Int. Conf. Solid-State Sensors and Actuators (Transducers-87)*, 1987, pp. 2-5.

[8] O. Brand and H. Baltes, "Micromachined resonant sensors - an overview," *Sensors update,* vol. 4, pp. 3-51, 1998.

[9] X. L. Feng, C. J. White, A. Hajimiri, and M. L. Roukes, "A self-sustaining ultrahigh-frequency nanoelectromechanical oscillator," *Nature nanotechnology,* vol. 3, pp. 342-346, 2008.

[10] D. B. Lesson, "A simple model of feedback oscillator noise spectrum," P*roc. IEEE,* vol. 54, pp. 329-330, 1966.

[11] E. Rubiola, "Phase noise and frequency stability in oscillators," Cambridge University Press, 2008, 20010, 2012.

[12] M. Sansa, E. Sage, E. C. Bullard, M. Gély, T. Alava, E. Colinet, A. K. Naik, L. G. Villanueva, L. Duraffourg, M. L. Roukes, G. Jourdan & S. Hentz, "Frequency fluctuations in silicon nanoresonators," *Nature Nanotechnology*, vol. 11, pp. 552–558, 2016.

[13] D. K. Agrawal and A. A. Seshia, "An analytical formulation for phase noise in MEMS oscillators," IEEE transactions on ultrasonics, ferroelectrics, and frequency control, vol. 61(12), pp.1938-1952, 2014.

[14] D. S. Greywall, B. Yurke, P. A. Busch, A. N. Pargellis, and R. L. Willett, "Evading amplifier noise in nonlinear oscillators," *Physical review letters,* vol. 72, p. 2992, 1994.

[15] L. G. Villanueva, R. B. Karabalin, M. H. Matheny, E. Kenig, M. C. Cross, and M. L. Roukes, "A nanoscale parametric feedback oscillator," *Nano letters,* vol. 11, pp. 5054-5059, 2011.

[16] E. Kenig, M. C. Cross, R. Lifshitz, R. B. Karabalin, L. G. Villanueva, M. H. Matheny, and M. L. Roukes, "Passive phase noise cancellation scheme," *Physical review letters,* vol. 108, p. 264102, 2012.

[17] D. K. Agrawal, J. Woodhouse, and A. A. Seshia, "Observation of locked phase dynamics and enhanced frequency stability in synchronized micromechanical oscillators," *Physical review letters,* vol. 111, p. 084101, 2013.

[18] M. H. Matheny, M. Grau, L. G. Villanueva, R. B. Karabalin, M. C. Cross, and M. L. Roukes, "Phase synchronization of two anharmonic nanomechanical oscillators," *Physical review letters,* vol. 112, p. 014101, 2014.

[19] C. Chen, D. N. H. Zanette, J. R. Guest, D. A. Czaplewski, and D. Lopez, "Self-sustained micromechanical oscillator with linear feedback," *Physical review letters,* vol. 117, p. 017203, 2017.

[20] L. S. Cao, D. X. Qi, R. W. Peng, M. Wang, and P. Schmelcher, "Phononic Frequency Combs through Nonlinear Resonances," *Physical Review Letters,* vol. 112, p. 075505, 2014.

[21] A. Ganesan, C. Do, and A. A. Seshia, "Phononic Frequency Comb via Intrinsic Three-Wave Mixing," *Physical Review Letters* 118 (3), 033903 (2017).

[22] A. Ganesan, C. Do, and A. A. Seshia, "Towards N-mode parametric electromechanical resonances," arXiv preprint arXiv:1708.01660, 2017.

[23] A. Ganesan, C. Do, and A. A. Seshia, "Observation of three-mode parametric instability in a micromechanical resonator," *Applied Physics Letters* 109 (19), 193501 (2016).

[24] A. Ganesan, C. Do, and A. A. Seshia, "Phononic Frequency Comb via Three-Mode Parametric Three-Wave Mixing," arXiv:1704.08008 (2017).

[25] J. Pettine, M. Patrascu, D.M. Karabacak, M. Vandecasteele, V. Petrescu, S.H. Brongersma, M. Crego-Calama and C. Van Hoof, "Volatile detection system using piezoelectric micromechanical resonators interfaced by an oscillator readout," *Sensors and Actuators A: Physical*, vol. 189, pp. 496-503, 2013.

[26] K. K. Park, H. J. Lee, G. G. Yaralioglu, A. S. Ergun, Ö. Oralkan, M. Kupnik, C. F. Quate, and B. T. Khuri-Yakub, "Capacitive micromachined ultrasonic transducers for chemical detection in nitrogen," *Applied Physics Letters*, vol. 91, pp. 094102, 2007.

[27] D. Rugar and P. Grütter, "Mechanical parametric amplification and thermomechanical noise squeezing," *Physical Review Letters*, vol. 67(6), p. 699., 1991.

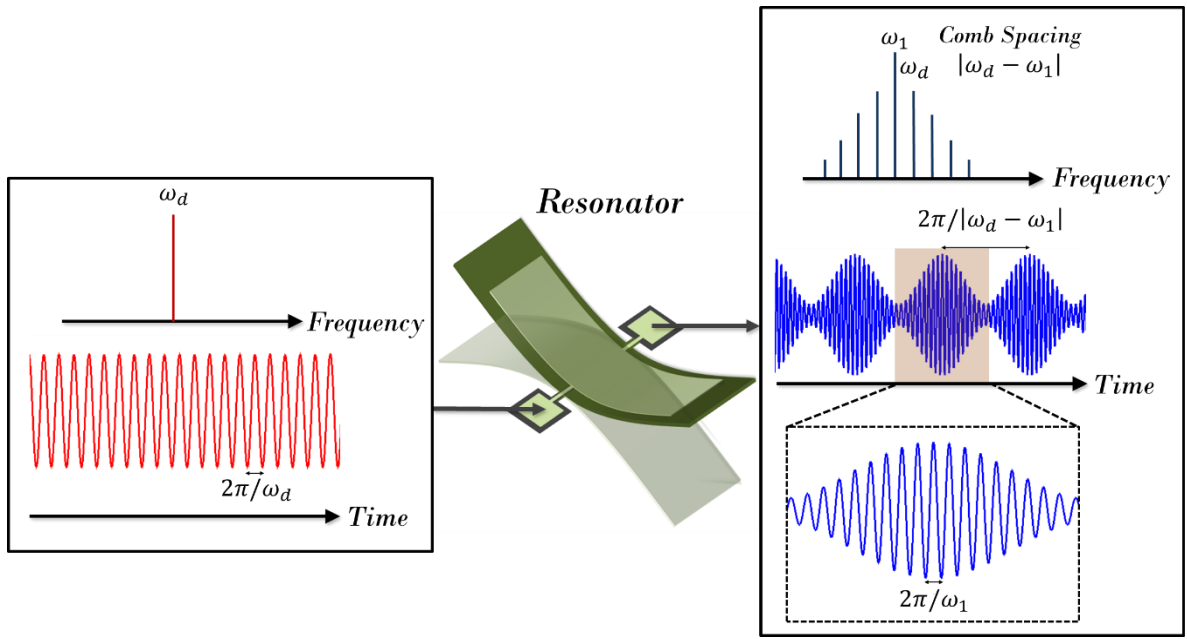

Figure 1: **Concept of resonant frequency tracking utilizing phononic frequency combs.** When a single sinusoidal drive is applied to a resonator, a series of equidistant spectral lines are generated through the phononic frequency comb generation process. In the time domain, the frequency comb corresponds to a mode-locked pulse train. In this frequency comb, the tone $\omega_1$ has the highest amplitude. This lays the basis for tracking resonant frequency $\omega_1$.

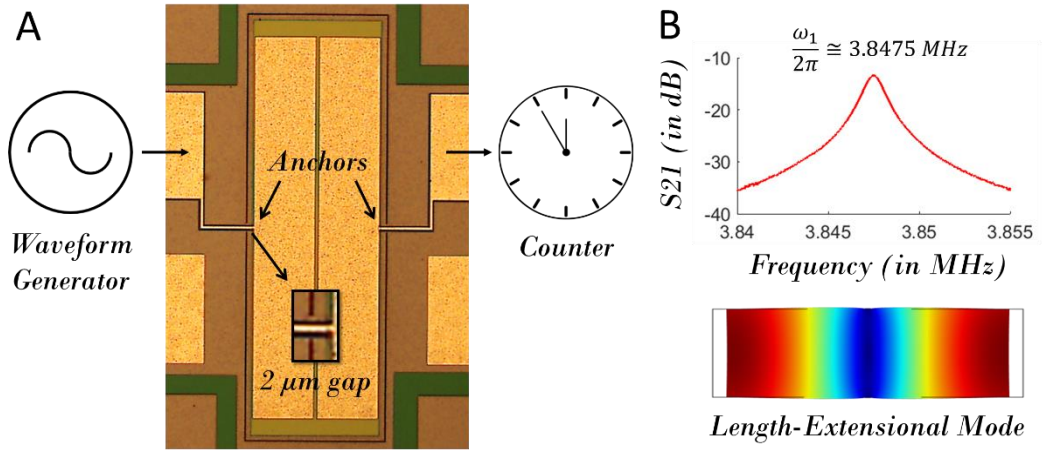

Figure 2: **Experimental setup for resonant tracking utilizing phononic frequency combs.** A: A drive signal derived from a waveform generator is fed to the micromechanical resonator and the output electrical signal is counted using a frequency counter; B: The resonant response of the length-extensional mode associated with the micromechanical resonator.

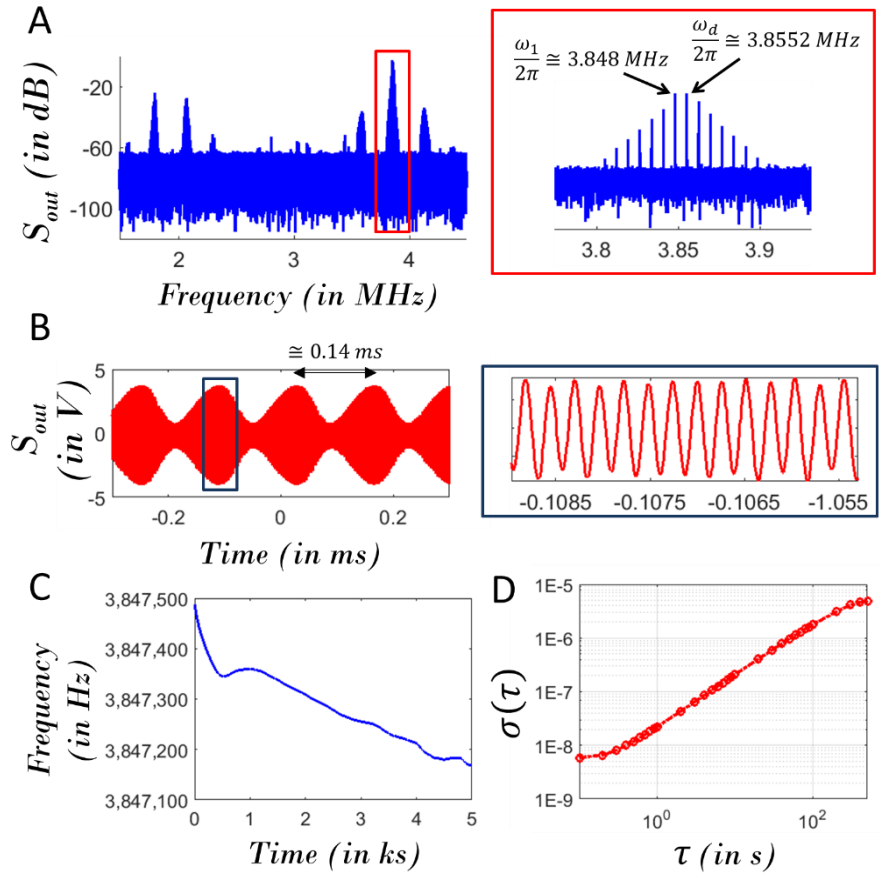

Figure 3: **First experimental demonstration of resonant tracking utilizing phononic frequency combs.** A: The frequency spectrum and B: Waveform corresponding to the output signal for the drive condition $S_{in}$(3.8552 MHz) = 10 dBm; C: The temporal evolution of the self-excited resonant frequency and D: the Allan Deviation $\sigma(\tau)$ of these frequency counts.

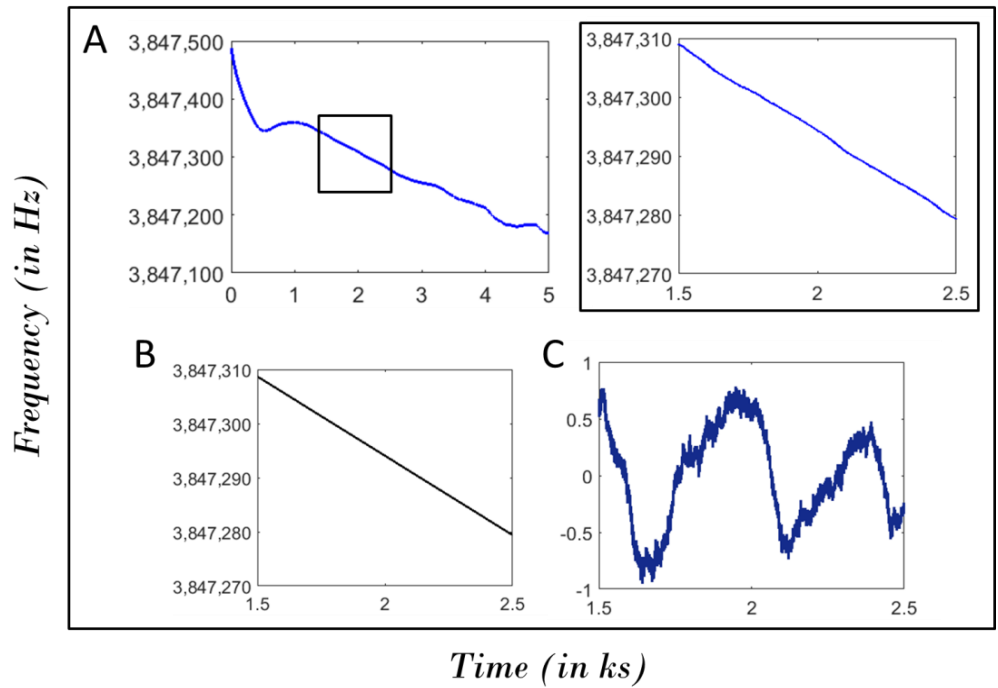

Figure 4: **Drift and fluctuations of resonant frequency.** A: The temporal evolution of self-excited resonant frequency and a small segment of this time series; B: The drift and C: fluctuations of the resonant frequency associated with this small segment.

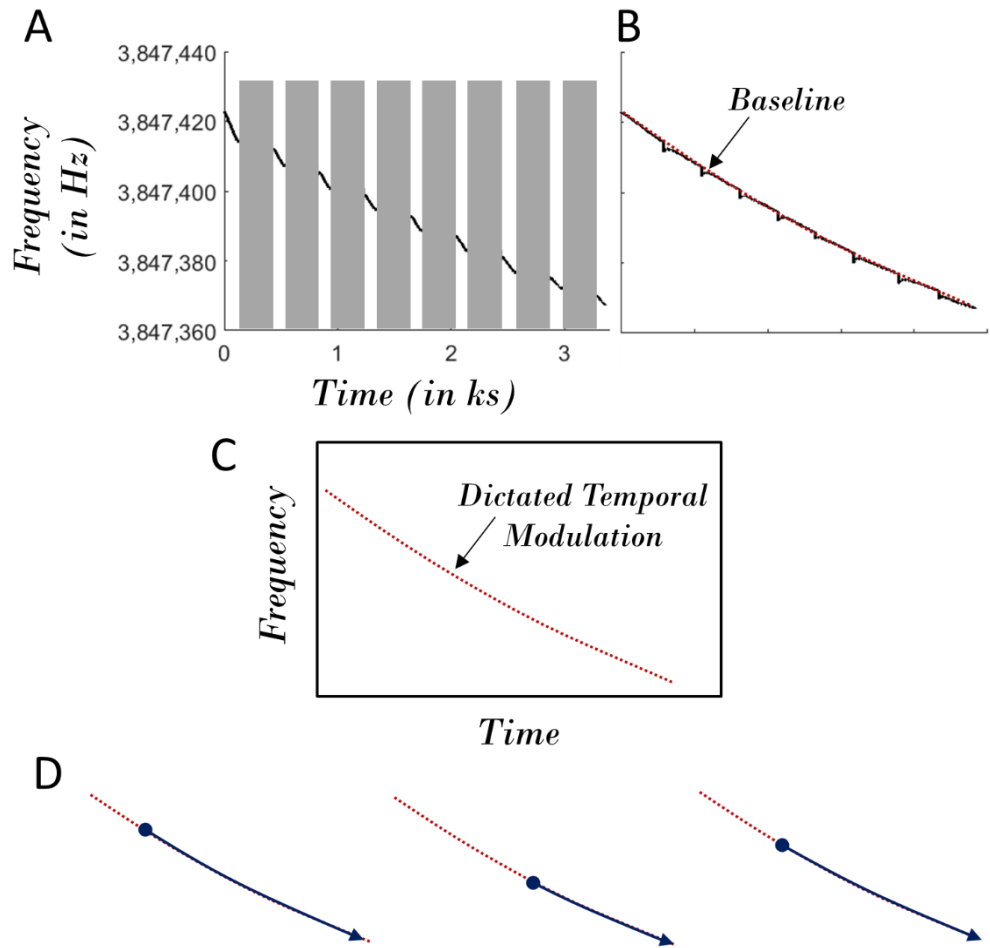

Figure 5: **Comb Induced Resonant Frequency Drift.** A: The temporal evolution of self-excited resonant frequency. The 'gray' regions correspond to the periods when the drive and hence the frequency comb process is switched-off; B: The concatenated frequency counts corresponding to the periods when the drive and hence the frequency comb process is switched-on; C: A characteristic temporal evolution profile associated with the self-excited resonant frequency which is dictated by the comb dynamics; D: The initial value dependent temporal evolution of self-excited resonant frequency. Note: This experiment was independently carried out to prove the comb induced resonant frequency drift. Hence, the associated dataset corresponding to this figure is not directly correlated to that associated with the other figures.

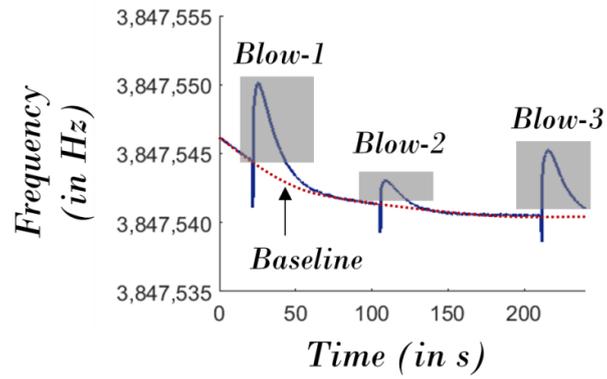

Figure 6: **Sensitivity of Self-Excited Resonant Frequency to Ambient Perturbations.** The temporal evolution of self-excited resonant frequency under ambient perturbations that are caused by external air flow. The 'gray' regions correspond to the response of resonant frequency to ambient perturbations. Note: This experiment was independently carried out to prove the sensitivity of self-excited resonant frequency to ambient perturbations. Hence, the associated dataset corresponding to this figure is not directly correlated to that associated with the other figures.